\documentclass[10pt]{article}
\usepackage{authblk}
\usepackage[a4paper,margin=1in]{geometry}
\usepackage{graphicx}
\usepackage{amsmath,amssymb}
\usepackage[hidelinks]{hyperref}
\usepackage[numbers,sort&compress]{natbib}

\begin{document}
\title{Strong Zeeman effects in the Fe I 5434.5 \AA{} line \\ observed in a sunspot on August 17, 2024}

\author[1]{Ivan I. Yakovkin}
\author[2]{Vsevolod G. Lozitsky}

\affil[1]{Institute of Physics of the National Academy of Sciences of Ukraine, Kyiv, Ukraine}
\affil[2]{Astronomical Observatory of the Taras Shevchenko National University of Kyiv, Kyiv, Ukraine}

\maketitle
\begin{abstract}
The Fe I 5434.5 \AA{} line is widely used as a non-magnetic reference due to its small effective Land\'e factor of $-0.014$. Nevertheless, this line can show significant non-zero Stokes $V$ polarization in active regions on the Sun. In this work, we study the distribution of magnetic fields in the sunspot observed on August 17, 2024 using the high-fidelity Stokes $I+V$ and $I-V$ spectra of three photospheric spectral lines of different magnetic sensitivity: Fe I 5434.5 \AA{}, Ni I 5435.9 \AA{} and Mn I 5432.5 \AA{}. The observed splitting of the Stokes $I\pm V$ bisectors in the Fe I 5434.5 \AA{} line follows the variation of the corresponding splitting in the two other spectral lines, indicating the presence of a spatial component with an extraordinarily strong magnetic field reaching 22 kG.

\end{abstract}

%


%
\section{Introduction}

Sunspot magnetic fields are among the most intensively studied phenomena in solar physics, since they provide direct insight into the structure of solar magnetism and are closely linked to energetic events such as solar flares and coronal mass ejections \citep{veronig2020can,cliver2022large}. Sunspot magnetic fields generally have a complex structure, which can be described by introducing multiple spatial components that differ in strength and filling factor \citep{rempel2012numerical}. The component with the largest filling factor is the most thoroughly investigated, as it produces clear and strong signatures of the Zeeman effect. Analyses of these signatures have shown that magnetic fields in mature sunspots are typically 2 -- 3 kG in magnitude, and can reach 4 -- 6 kG in exceptional cases \citep{solanki2003sunspots,livingston2006sunspots}. 

Meanwhile, the magnetic field strength in the components with small filling factors still remains a subject of active research. Some studies estimate that the magnetic field in the components with low filling factors can reach 7 -- 8 kG in magnitude \citep{van2013peripheral,lozitsky2016indications,okamoto2018super,duran2020detection}. 
Such estimates, however, are often less definitive, as they generally contribute to finer features in the observed spectra \citep{solanki1993small}. The signatures of Zeeman and Doppler effects become barely distinguishable from those in the more dominant components, while the signatures of strong Zeeman or Doppler effects in components with a small filling factor can be superimposed onto the spectra of the adjacent umbral and molecular lines or be significantly attenuated due to spatial cancellation \citep{bellot2019quiet}. It should also be noted that since the dominant components generally have a lower magnetic field strength, many instruments and inversion codes are developed under the assumption of weaker magnetic fields and are therefore not optimized for detecting much stronger fields \citep{del2016inversion}. Nevertheless, a recent study of Hinode/SOT data has revealed that superstrong magnetic fields exceeding 8 kG are in fact rather common in the sunspots of certain topologies \citep{duran2025superstrong}.

One possible way to detect strong Zeeman effects can lie in using spectral lines with very low Land\'e factors for the magnetic field measurements. While such an approach is inherently less precise, a low Land\'e factor ensures that the spectral features are located closer to the line core, facilitating their identification. In this regard, the analysis of bisectors of the Stokes $I\pm V$ profiles is a powerful diagnostic for identifying the spatial inhomogeneity of the magnetic field \citep{lozitsky2015small}.  A magnetic parameter $B^*$, derived from the bisector splitting $\Delta\lambda_{bis}$ as:
\begin{equation}\label{bstar}
  B^*= \dfrac{\Delta\lambda_{bis}/2}{4.67\times10^{-13}g_{eff}\lambda^{2}},   
\end{equation}
is particularly useful for quick estimates of the magnetic field: in the case of a uniform magnetic field, the bisectors of the Stokes $I\pm V$ profiles are mutually parallel for weak or purely longitudinal magnetic fields with $B^*\approx B_\parallel$, while the bisector splitting near the line core is related to the modulus of the magnetic field: $B^*\approx |B|$. In Eq. (\ref{bstar}), $\Delta\lambda_{bis}$ is the full splitting of the bisectors, $g_{eff}$ is the effective Land\'e factor, and $\lambda$ is the wavelength of the spectral line under consideration. In this work, we use multiple spectral lines with significantly different Land\'e factors to simultaneously capture both weak and strong magnetic components. In particular, the Fe I 5434.5 \AA{} line was included in the analysis as it has a very small effective Land\'e factor of $-0.014$ \citep{landi1982effective}, which explains its wide use as a non-magnetic diagnostic \citep{sankarasubramanian2003properties,wunnenberg2002evidence}.

\section{Observations and data processing}

\begin{figure}
	\centerline{\includegraphics[width=\textwidth]{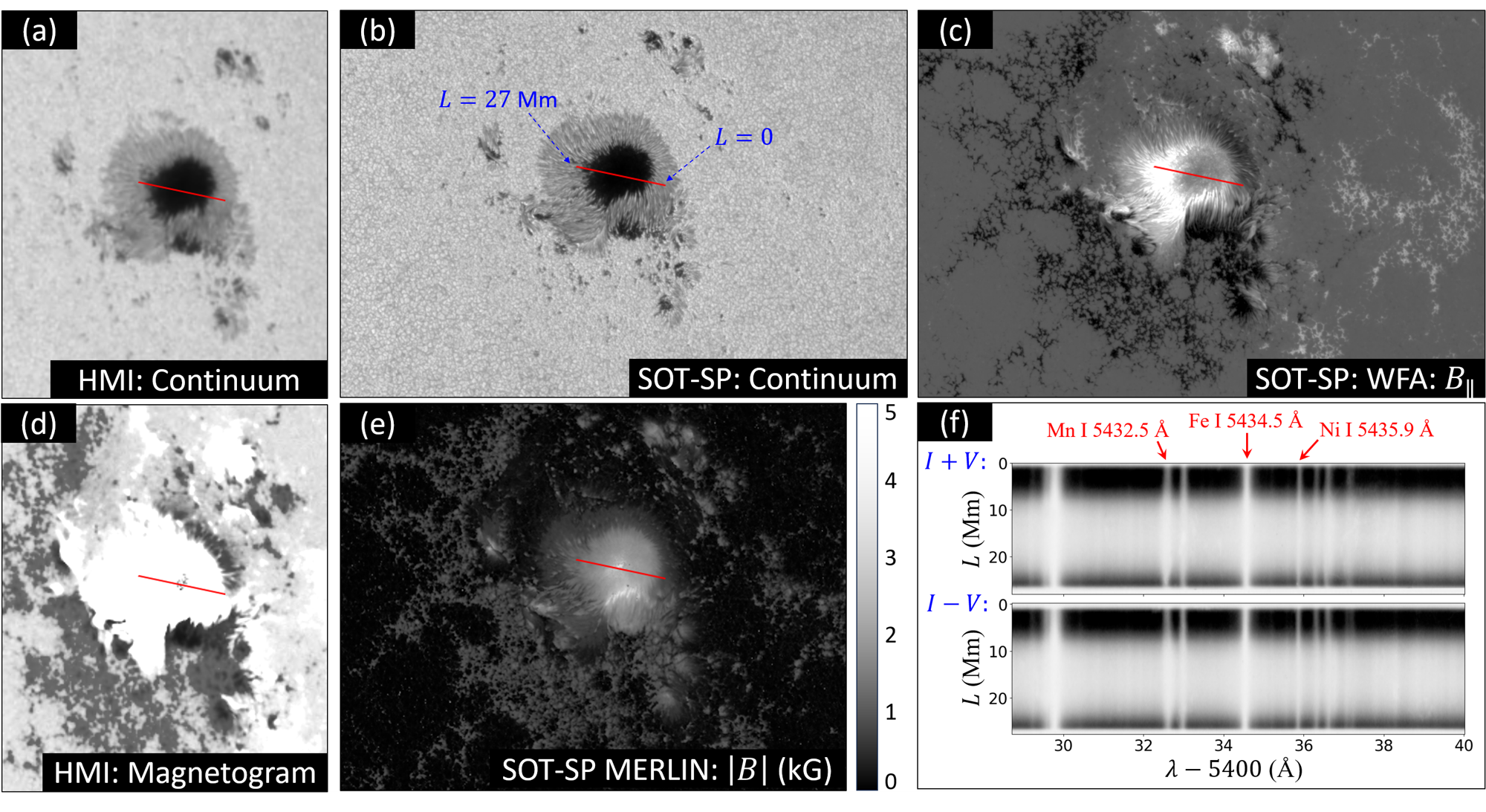}}
    \caption{Sunspot on 2024 August 17. SDO/HMI: continuum map (a) and magnetogram (d); Hinode/SOT-SP: continuum map (b), line-of-sight magnetic field in the weak-field approximation (c), MERLIN-inverted magnetic field magnitude (e); Stokes $I\pm V$ spectra observed with HST AO KNU in the $5429-5440$ \AA{} range (f). The spectrograph slit is marked with a red interval.}
    \label{fig:spot}
\end{figure}

In this study, we investigate the largest sunspot in active region NOAA 3784, which was observed on 2024 August 17. Observations were carried out with the Echelle spectrograph of the Horizontal Solar Telescope of the Astronomical Observatory of Taras Shevchenko National University of Kyiv (HST AO KNU, see Appendix \ref{hst}). This spot was located at $\mu=\cos{\theta}=0.845$, corresponding to a heliocentric angle of $\theta \approx 32^\circ$ from the disk center and was approximately $40$ Mm in size (Fig. \ref{fig:spot}a,b). The magnetic configuration of the group was of the $\beta\gamma\delta$ type. According to visual measurements made in the Fe I 5250.2 \AA{} line, the magnetic field strength in it reached 3700 G with the N polarity. 

Figure \ref{fig:spot}a shows the SDO/HMI continuum map of the sunspot at 06:55:21 UT, and Fig. \ref{fig:spot}d shows the corresponding magnetogram. Figure \ref{fig:spot}c shows the distribution of the longitudinal magnetic field at 11:19:39 UT, calculated within the weak-field approximation from the Hinode SOT/SP data \citep{kosugi2007hinode,lites2013hinode}, and Fig. \ref{fig:spot}e shows the corresponding magnitude of the magnetic field obtained through MERLIN inversion within the Milne-Eddington approximation. 

The spectrum of this sunspot was recorded with an exposure time of 25 s, starting at 6:55:00 UT using the WP3 ORWO photoplates. The slit of the spectrograph crossed the sunspot through its center in the east-west direction, as marked by the red interval in Fig. \ref{fig:spot}a--e. The equivalent length of the spectrograph slit was approximately 27 Mm, capturing the penumbra and umbra of the spot. A portion of the obtained spectrogram covering the $5429-5440$ \AA{} spectral region is shown in Fig. \ref{fig:spot}f, where the upper and lower panels correspond to Stokes $I-V$ and $I+V$ signals, respectively. Here, darker colors correspond to higher Stokes $I\pm V$ intensity. In this article, we use the convention for the Stokes $V$ sign that matches \cite{landi2004polarization}: positive Stokes $V$ corresponds to right-handed circular polarization (clockwise rotation of the electric field vector as seen by an observer facing the source).

\section{Results}

\begin{figure}
	\centerline{\includegraphics[width=0.9\textwidth]{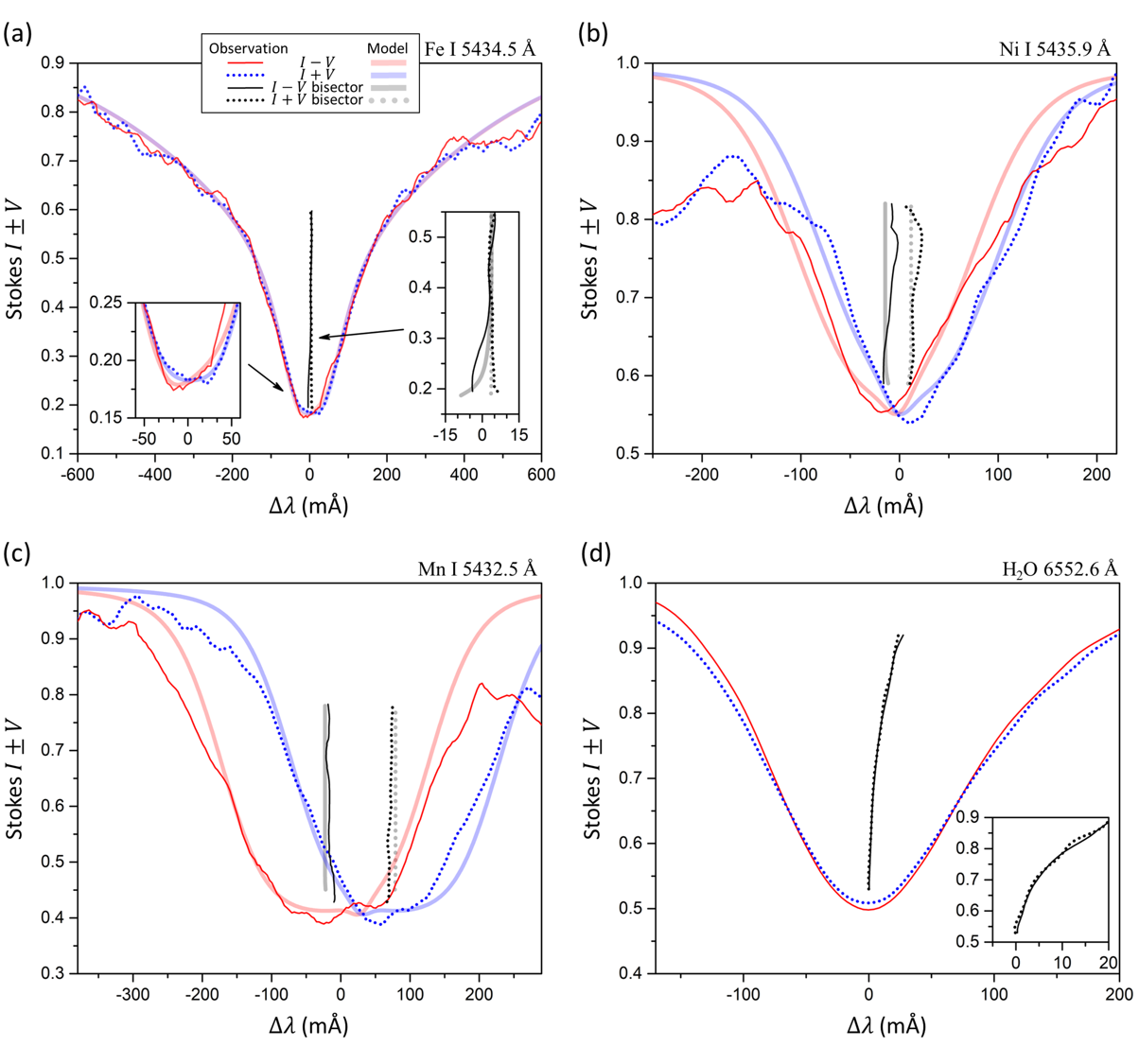}}
\caption{Stokes $I\pm V$ profiles and bisectors of the Fe I 5434.5 \AA{} (a),  Ni I 5435.9 \AA{} (b), Mn I 5432.5 \AA{} (c), and H$_2$O 6552.6 \AA{} (d) lines corresponding to $L=21$ Mm. Insets in panels (a) and (d) show enlarged views of the profiles and bisectors.}
    \label{fig:ipmv}
\end{figure}

Figure \ref{fig:ipmv} shows the Stokes $I\pm V$ profiles of the three lines under study (Fe I 5434.5 \AA{}, Ni I 5435.9 \AA{} and Mn I 5432.5 \AA{}) and the telluric H$_2$O 6552.6 \AA{} line at the location $L=21$ Mm along the slit (see Fig. \ref{fig:spot}b). Each panel also shows the bisectors (centroids) of the absorption peaks in each of the two circular polarizations to highlight any relative shifts between the $I+V$ and $I-V$ profiles. The insets in Fig. \ref{fig:ipmv}a,d show enlarged views of the profiles and bisectors. Thicker semi-transparent lines correspond to the model calculations, as will be detailed below. The effective Land\'e factors of the lines under study are $-0.014$ \citep{landi1982effective},  $0.500$ \citep{litzen1993spectrum}, and $2.143$ \citep{sugar1984atomic} for the Fe I 5434.5 \AA{}, Ni I 5435.9 \AA{}, and Mn I 5432.5 \AA{} lines, respectively. The wavelength registration of the  Stokes $I+V$ and $I-V$ was performed using the telluric H$_2$O 6552.6 \AA{} line (Fig. \ref{fig:ipmv}d), which, in contrast to the Fe I 5434.5 \AA{} line, shows essentially perfect alignment of bisectors both in the line core and in its wings. 

The bisectors of the Ni I 5435.9 \AA{} and Mn I 5432.5 \AA{} lines are mostly vertical and mutually parallel in $I\pm V$ polarizations (Fig. \ref{fig:ipmv}b, c), with the splitting corresponding to the magnetic parameter $B^*$ of approximately $2$ kG in both lines. Meanwhile, the bisectors of the Fe I 5434.5 \AA{} line diverge near the line core (Fig. \ref{fig:ipmv}a), well above the estimated measurement uncertainties with a corresponding value of $B^*$ of approximately $-26$ kG.

\begin{figure}
	\centerline{\includegraphics[width=0.8\textwidth]{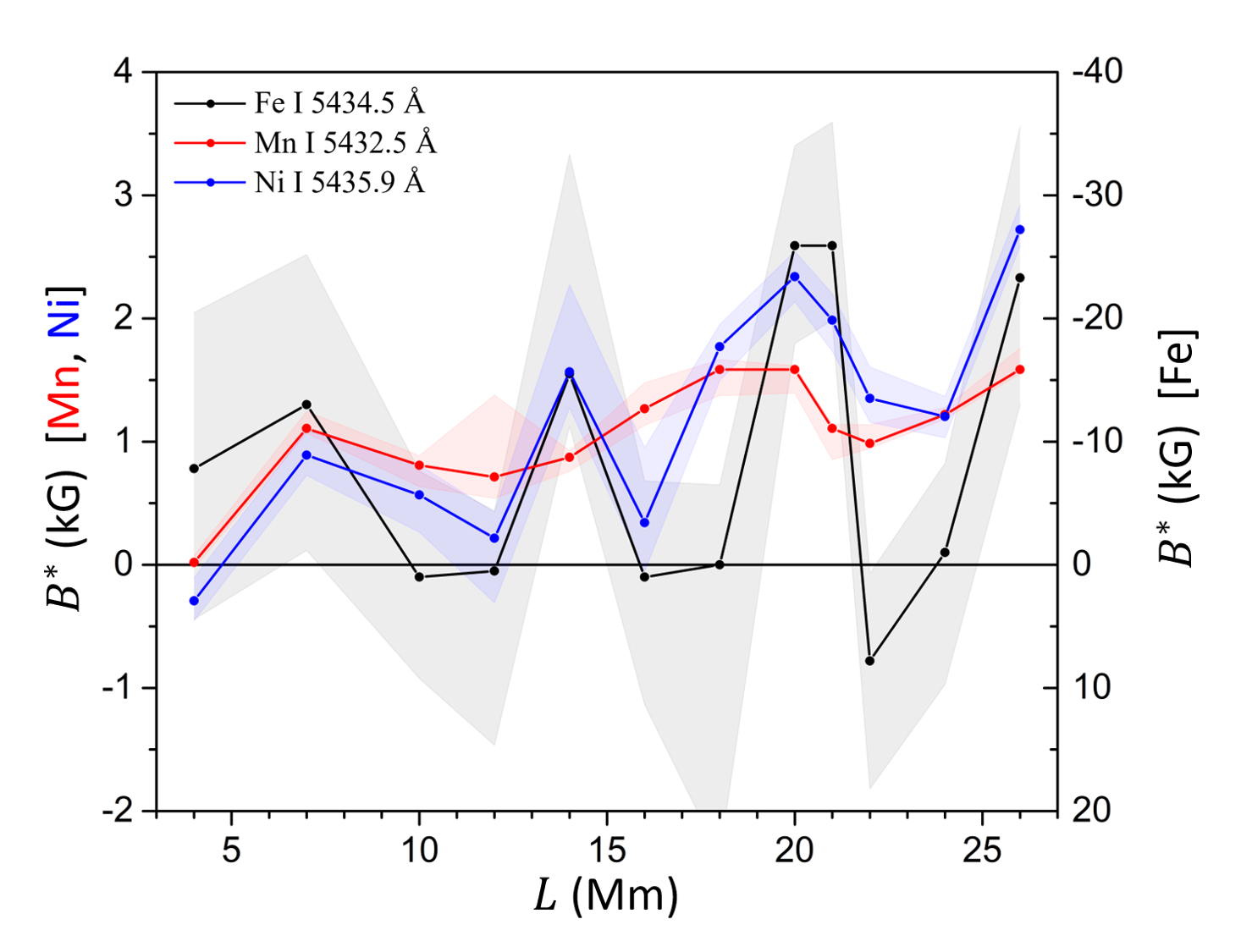}}
\caption{
Comparison of the magnetic parameter $B^*$ in the sunspot on August 17, 2024, measured in the Fe I 5434.5 \AA{},  Ni I 5435.9, and Mn I 5432.5 \AA{} lines at different locations $L$ along the spectrograph slit (Fig. \ref{fig:spot}b). The values of $B^*$ for the Ni I 5435.9 and Mn I 5432.5 \AA{} lines correspond to the left scale, while those for the Fe I 5434.5 \AA{} line correspond to the right scale. 
}
    \label{fig:mag}
\end{figure}

The distribution of the magnetic field parameter $B^*$ along the spectrograph slit is shown in Fig. \ref{fig:mag}, where $B^*$ of the Fe I 5434.5 \AA{} line uses the scale on the right to facilitate comparison. The shaded areas denote the asymmetric $1\sigma$ confidence band (see Appendix \ref{errors} for details). The variation of the magnetic field along the slit indicates a fine stratification of the photospheric magnetic fields. At the same time, the measured magnetic parameter $B^*$ in the  Fe I 5434.5 \AA{} line is an order of magnitude larger than the corresponding measurements in the other two spectral lines, reaching approximately 25 kG at $L\approx20-21$ Mm. It should be emphasized that the opposite sign of the magnetic parameter $B^*$ in the Fe I 5434.5 \AA{} line does not imply an opposite polarity of the underlying magnetic field, as will be shown below. This sign reversal reflects the absorption-emission interplay inferred from the model: the strongly magnetized upper component contributes mainly as an emissive feature above lower, optically thicker absorptive components. The location with maximum measured magnetic field parameter $B^*$ corresponds to the eastern part of the sunspot, where the large-scale magnetic field of the spot can be expected to be mostly longitudinal (see Fig. \ref{fig:spot}c,e).

While the Fe I 5434.5 \AA{} line indicates some extremely strong magnetic field values exceeding 20 kG in some locations $L$ along the spectrograph slit, the changes in the magnetic fields along the slit are well aligned in all three spectral lines. The Spearman rank-order correlations across 11 slit positions between Ni I 5435.9 \AA{} and Mn I 5432.5 \AA{} lines ($r_s = 0.72$, $p = 0.009$) and between Fe I 5434.5 \AA{} and  Ni I 5435.9 \AA{} lines ($r_s = 0.61$, $p = 0.035$) indicate positive associations, while the correlation between Mn I 5432.5 \AA{} and Fe I 5434.5 \AA{} lines ($r_s = 0.32$, $p = 0.31$) was positive but not statistically significant. This result is consistent with the formation heights of these lines: the Mn I 5432.5 \AA{} line forms the lowest, and the Fe I 5434.5 \AA{} line forms the highest in the solar photosphere \citep{wiehr1996spatial,degenhardt1994evershed, lohner2019convective, vitas2005heights, grossmann1994height}, which explains why Fe I 5434.5 \AA{} and Mn I 5432.5 \AA{} lines show the weakest association. Overall, the Mn I 5432.5 \AA{} line shows a more gradual variation of magnetic parameter $B^*$ along the slit. This line is formed by a spin-changing intercombination transition, and it is therefore forbidden in strict LS coupling, giving it a transition probability several orders of magnitude lower than the allowed Fe I 5434.5 \AA{} and  Ni I 5435.9 \AA{} lines and decreasing its formation height.

A possible explanation for the difference in the magnetic fields measured in the Fe I 5434.5 \AA{}, Ni I 5435.9 \AA{}, and Mn I 5432.5 \AA{} lines can be illustrated by a plane-parallel model atmosphere consisting of three components (Table \ref{tbl:solrat}). We note that this model is not intended to be a unique inversion of the sunspot atmosphere. It is a minimal forward model showing how line formation in absorptive and emissive components with the same magnetic polarity can produce a bisector splitting with an apparent sign reversal of the  magnetic parameter $B^*$ measured in the Fe I 5434.5 \AA{} line. The lowest component (I) has a magnetic field $B=1.9$ kG, and captures the formation of the Ni and Mn lines. Due to the longitudinal orientation of the field, the mutually parallel bisectors in Ni and Mn lines capture the longitudinal magnetic field in this component: $B^* \approx B_\parallel=1.9$ kG. The component in the middle (II), located roughly at the temperature minimum zone, is responsible for the formation of the Fe line core. It should be emphasized that component (II) is assumed to be non-magnetic only as a simplifying approximation to reduce the number of fitting parameters; introducing a weak magnetic field in this layer does not qualitatively affect the inferred properties of component (III). The highest component (III) with a small optical depth $\Delta\tau$ carries the magnetic field of $22$ kG of the same polarity as component (I). The calculations of the Stokes profiles were performed within the local thermodynamic equilibrium (LTE) approximation using the forward-modeling SolRaT code \citep{Yakovkin2023SolRaT} (Fig. \ref{fig:ipmv}a--c, semi-transparent lines; for details see Appendix \ref{solrat}). Due to the larger formation height of the Fe I 5434.5 \AA{} line, it is much more sensitive to the component (III) compared to the Ni I 5435.9 \AA{} and Mn I 5432.5 \AA{} lines, and the bisector splitting near the line core captures the magnitude of the magnetic field $|B^*|\approx|B|=22$ kG. It should also be noted that while the magnetic parameters $B^*$ for the Fe I 5434.5 \AA{} line are of the opposite sign compared to the Ni I 5435.9 \AA{} and Mn I 5432.5 \AA{} lines, the actual magnetic field has the same sign in all three lines due to the mostly absorptive contribution of components (I) and (II), and a more emissive contribution from component (III).

\section{Discussion}

The results presented in Fig. \ref{fig:mag} show that the magnetic fields measured in the Ni I 5435.9 \AA{} and Mn I 5432.5 \AA{} lines reach $2-3$ kG in magnitude. The order of magnitude of these magnetic fields is in agreement with the magnetic fields obtained through automatic MERLIN inversion of Hinode SOT/SP data (Fig. \ref{fig:spot}e). Such magnetic fields correspond to the $\Delta \lambda_{H}\sim \Delta\lambda_{1/2}$ regime in the Ni I 5435.9 \AA{} and Mn I 5432.5 \AA{} lines, which, together with the apparently mutually parallel bisectors in these lines (Fig. \ref{fig:ipmv}b,c), indicate that the locations with $B\gtrsim 1$ kG in Fig. \ref{fig:mag} correspond to predominantly longitudinal magnetic fields. The latter is in good agreement with the longitudinal magnetic field calculated within the weak-field approximation from Hinode SOT/SP data (Fig. \ref{fig:spot}c), which shows the increased longitudinal magnetic field of  $\gtrsim1$ kG around $L\approx 6$ Mm and $L\gtrsim21$ Mm. 

Meanwhile, the Fe I 5434.5 \AA{} line shows reliable splitting of the Stokes $I\pm V$ bisectors in the line core, corresponding to magnetic fields larger by an order of magnitude and with an opposite polarity compared to the Ni I 5435.9 \AA{} and Mn I 5432.5 \AA{} lines. This raises the question of whether such splittings can be interpreted as a manifestation of the Zeeman effect and whether such extraordinarily large magnetic fields exceeding 20 kG can exist in sunspots. It is worth recalling that the possibility of the magnetic fields in sunspots reaching $10^4$ G was first expressed in \citet{severny1957some}. 
Later, in a follow-up study, an important result was obtained by comparing measurements of solar magnetic fields outside the spots using 11 spectral lines with Land\'e factors ranging from 1 to 3: it was concluded that the lines with smaller Land\'e factors indicate larger longitudinal magnetic fields when measured with a magnetograph \citep{gopasyuk1973comparison}. To explain this relation, a two-component model of the magnetic field was proposed, in which the spatially unresolved small-scale component had local magnetic field strengths reaching 10 kG \citep{lozitsky1980calibration}.
Regarding the Fe I 5434.5 \AA{} line specifically, the magnetic field measurements in solar flares using this line reached $10^4$ G \citep{lozitskij1998observations}, with a magnetic field stratification similar to that inferred in this work for the sunspot on August 17, 2024. While, to our knowledge, no other instruments that observed this sunspot indicated the magnetic fields of such magnitude, this feature could have been rather short-lived due to pressure imbalance, making them harder to detect. Additionally, it should be noted that many instruments are designed to operate with magnetic fields of only a few kG in magnitude, and therefore can miss the manifestation of such fields.

The obtained data can be interpreted within a three-component structure of the sunspot magnetic field, which contains two dominant components (I) and (II) with the magnetic fields of $0-3$ kG and a component (III) with a much smaller optical depth $\Delta\tau$ but a much stronger magnetic field in the $10^4$ G range. Within this framework, the Ni I 5435.9 \AA{} and Mn I 5432.5 \AA{} lines probe the $0-3$ kG component in the  $\Delta \lambda_{H}\sim \Delta\lambda_{1/2}$ regime, mostly reflecting its longitudinal magnetic field, while the Fe I 5434.5 \AA{} line probes the $10^4$ G component within the  $\Delta \lambda_{H}\ll \Delta\lambda_{1/2}$ regime. 

This in turn raises a question whether magnetohydrodynamic stability is possible at the photospheric level at such high magnetic field magnitudes. Indeed, high magnetic fields are believed to become unstable due to pressure balance, meaning that any features with high magnetic fields can only be transient. However, Solov'ev \cite{solov2022force} recently proposed a theoretical model for twisted force-free magnetic flux tubes that can support magnetic fields of such magnitude. Additionally, the observations presented in this work can relate to transient features, which significantly relaxes the limitations imposed by the force-free pressure balance condition.

This interpretation also explains why the unusually large value inferred from the Fe I 5434.5 \AA{} line is not recovered by most other satellite or ground-based measurements. The visual measurement of about 3700 G in Fe I 5250.2 \AA{} and the Hinode SOT/SP values shown in Fig. \ref{fig:spot}c,e are consistent with the expected field strength of the dominant umbral component, and our Ni I 5435.9 \AA{} and Mn I 5432.5 \AA{} measurements sample the same component. They therefore provide an important consistency check rather than a contradiction. By contrast, the Fe I 5434.5 \AA{} diagnostic used here is sensitive to a weak, spatially unresolved component that contributes only a small fraction of the total optical depth but produces a measurable core splitting because the line remains in the $\Delta \lambda_{H}\ll\Delta\lambda_{1/2}$ regime even for very strong fields. Standard magnetograms and Milne-Eddington inversions are generally dominated by the larger-filling-factor atmosphere, use lines with substantially larger Land\'e factors whose strong-field signatures can be saturated, blended, or diluted by stray light and spatial averaging, and often assume one or a small number of magnetic components. As a result, they preferentially return the $ 3.5$ kG field of the main sunspot atmosphere and are not expected to isolate the small-filling-factor component responsible for the bisector splitting of the Fe I 5434.5 \AA{} line.

A separate note should be made on the value of the effective Land\'e factor of the Fe I 5434.5 \AA{} line. The theoretical Land\'e factor for this line within the LS coupling approximation is exactly zero, while the experimental measurements report the values of $g_{eff}$ from $-0.010$ \citep{kochukhov2020hidden} to $-0.014$ \citep{landi1982effective}. Here, it should be noted that most theoretical estimates of the effective Land\'e factors correspond to the linear Zeeman splitting regime. However, in stronger magnetic fields $g_{eff}$ becomes dependent on the magnetic field $B$ due to the incomplete Paschen-Back effect. This nonlinear behavior necessitates the use of the field-dependent effective Land\'e factors $g_{eff}(B)$, calibrated specifically for the magnetic field strengths of interest \citep{yakovkin2024altitude}. While, to our knowledge, no significant increase in the magnetic sensitivity of the Fe I 5434.5 \AA{} has been reported, the potential of $g_{eff}$ deviating from the $-0.014$ value is a topic for future research. While the measurements in the lines with such low Land\'e factors are inherently less precise, it can be argued that such lines can be the only ones that can detect magnetic fields of such magnitude: if a line with a large magnetic sensitivity is observed under especially strong magnetic fields, the Zeeman components are likely to be lost among the adjacent lines, and the attribution of residual Stokes $V$ polarization to the correct spectral line becomes unreliable.

Non-LTE effects could also contribute to the observed splitting of Stokes $I\pm V$ profiles. Atomic level polarization (ALP) can, in general, result in a non-zero Stokes $V$ signal even in the absence of a magnetic field. In addition to non-LTE effects, fine atmospheric structuring can lead to the Stokes $I\pm V$ profiles resembling significantly stronger magnetic fields \citep{yakovkin2024altitude}. The interplay between the magnetic fields and the ALP (mainly through the Hanle effect) is widely used for measuring the magnetic fields of lower magnitude \citep{shchukina2011determining,milic2012hanle}. Nevertheless, the high coherence relaxation rates due to collisions in the photosphere significantly reduce the ALP \citep{landi2004polarization}, drastically reducing the impact of the non-LTE effects in photospheric lines. Therefore, while non-LTE effects in photospheric lines generally contribute to the line strength, formation height, and line shape, the direct influence on the Stokes $V$ signal is generally weaker \citep{holzreuter2015three, collet2005effects,smitha2023non,smitha2020influence}. A similar consideration applies to dynamic processes such as acoustic waves, atmospheric gravity waves, and unresolved hot or cool streams. These processes can produce Doppler shifts, broadenings, and asymmetries in the Fe I 5434.5 \AA{} intensity profile, as shown in previous non-magnetic applications of this line \citep{andjic2007energy,kaisig1982asymmetry,kneer2011acoustic}. While these processes do not introduce a Stokes $V$ signal directly, they may affect the line formation under non-LTE conditions when combined with other effects.

The possible influence of Stark shifts and broadening should also be considered. Generally, the Stark effect most directly manifests as broadening and splitting in 
Stokes $I$ and as linear polarization signatures in Stokes $Q$ and $U$. Since our 
analysis is based on Stokes $I \pm V$, the Stark effect enters only at second order, 
through mechanisms such as indirect coupling between linear and circular polarization 
via magneto-optical effects and alignment-to-orientation conversion. While these second-order effects generally do not produce a large enough anti-symmetric Stokes $V$ signal, a separate detailed consideration of possible manifestations of the Stark effect in the Fe I 5434.5 \AA{} line is a topic for future work.

\section{Conclusions}

In this work, we investigated the magnetic field structure of the sunspot in active region NOAA 3784 observed on 17 August 2024, using high-fidelity Stokes $I+V$ and $I-V$ spectra of three photospheric lines with substantially different magnetic sensitivities: Fe I 5434.5 \AA{}, Ni I 5435.9 \AA{}, and Mn I 5432.5 \AA{}. The Ni I 5435.9 \AA{} and Mn I 5432.5 \AA{} lines indicate magnetic field strengths of approximately $2-3$ kG along the spectrograph slit, in agreement with independent Hinode SOT/SP measurements and typical values for mature sunspots. The bisector behavior of these two lines suggests that the magnetic field in the dominant photospheric component (I) is predominantly longitudinal in the umbra.

At the same time, the Fe I 5434.5 \AA{} line, despite its extremely small catalog effective Land\'e factor, shows a clear splitting of the Stokes $I\pm V$ bisectors in the line core at multiple locations along the slit, with a spatial variation that closely follows that observed in the Ni I 5435.9 \AA{} and Mn I 5432.5 \AA{} lines. Interpreted within the three-component atmosphere model, this splitting implies extraordinarily strong magnetic fields up to 22 kG. The  tight spatial correlation of magnetic field variations inferred from the Fe I 5434.5 \AA{} and Ni I 5435.9 \AA{} lines demonstrates that the Fe I 5434.5 \AA{} line, traditionally regarded as nearly non-magnetic, may be sensitive to underlying unresolved photospheric magnetic fields in sunspots. 
\\
\\
\noindent\textbf{Acknowledgements:}
Hinode is a Japanese mission developed and launched by ISAS/JAXA, with NAOJ as domestic partner and NASA and STFC (UK) as international partners. It is operated by these agencies in co-operation with ESA and NSC (Norway). SDO/HMI data is courtesy of NASA/SDO and the HMI science teams.

This research was funded by the Ministry of Education and Science of Ukraine (Kyiv, UA) grant numbers 22BF023--03 and 25BF051--04.

\appendix   

\section{Observations on HST AO KNU}
\label{hst}
The optical scheme of the telescope and instrumental details are described in \citet{lozitsky2016indications}. 
The main advantage of the observations made on the Echelle spectrograph is that they cover a vast spectral range from 3800 to 6600 \AA{}, in which Stokes  $I+V$ and $I-V$ signals are recorded simultaneously on the adjacent bands of the spectrograms. This ensures that the obtained Stokes  $I+V$ and $I-V$ spectra capture the same location at the same point in time, which is critical for subsequent magnetic field measurements. The spectrograms were digitized using the Epson Perfection V 550 high-resolution scanner and then underwent the standard pre-processing procedure detailed in \citet{yakovkin2023search}. The sampling spectral resolution was approximately $3.5$ m\AA{}, with the registration uncertainty within 5 m\AA{}. While the registration was performed in the H$_2$O 6552.6 \AA{} line, the registration accuracy on HST AO KNU is consistent across large spectral intervals including the vicinity of the Fe I 5434.5 \AA{} line.

\section{Measurement errors and uncertainties}
\label{errors}

To capture the asymmetry of the magnetic field measurement errors arising from the specific shapes of the $I \pm V$ profiles, the uncertainty of the magnetic parameter $B^*$ was estimated from the uncertainties of the $I \pm V$ profiles using Monte Carlo sampling. The shaded region in Fig. \ref{fig:mag} corresponds to an interval $[B_-, B_+]$ defined as
\begin{equation}
    \int_{-\infty}^{B_-} p(B)\,\mathrm{d}B = \int_{B_+}^{\infty} p(B)\,\mathrm{d}B = \dfrac{1}{2}\left(1-\operatorname{erf}\!\left(\frac{1}{\sqrt{2}}\right)\right),
\end{equation}
where $p(B)$ is the probability density function of the inferred magnetic field, $\operatorname{erf}$ is the Gauss error function, and $B_-$ and $B_+$ denote the lower and upper bounds of the confidence band, respectively. By construction, each tail of $p(B)$ lying outside the interval carries a probability of $\tfrac{1}{2}\left(1-\operatorname{erf}\!\left(1/\sqrt{2}\right)\right)
\approx 0.1587$, so that $[B_-, B_+]$ contains ${\approx}\,68.27\%$ of the total
probability. For a Gaussian $p(B)$ this equal-tailed interval reduces exactly to $[\mu-\sigma,\,\mu+\sigma]$; for the asymmetric distributions obtained here, $B_-$ and $B_+$ are in general not equidistant from the adopted value $B^*$. The bounds are additionally capped as $B_-\!\le\!B^*\!\le\!B_+$, so that the
shaded band always encloses $B^*$.

\section{Forward modeling details}
\label{solrat}

Forward modeling was performed using the SolRaT code \citep{Yakovkin2023SolRaT}, in which the radiative transfer problem for the plane-parallel atmosphere was solved within the multi-term atom formalism, assuming an LTE solution of statistical equilibrium equations \citep{landi2004polarization}. The atmosphere was modeled with three vertically stacked horizontal constant-property slabs (components). Component (I) represents the ordinary umbral photosphere and dominates the Ni I and Mn I line formation. Component (II) is a cool layer with large Fe I opacity that mainly shapes the Fe I 5434.5 \AA{} absorption core. Component (III) is optically thin and contributes mainly as a narrow emissive feature formed above these stronger absorptive layers, while carrying the very strong field required to reproduce the Fe I core splitting. Each slab was described by the temperature ($T$), optical thicknesses of the continuum ($\Delta \tau_{continuum}$) and the line center ($\Delta\tau_{Fe}$/$\Delta\tau_{Ni}$/$\Delta\tau_{Mn}$), the magnetic field magnitude ($B$), the two angles ($\chi_{B}$ and $\theta_{B}$) defining the direction of the latter, the Voigt profile damping coefficient $a$, the turbulent velocity $v_{turb}$, and the macroscopic velocity $v_{mac}$ (Table \ref{tbl:solrat}). 

\begin{table}
    \caption{Atmosphere parameters used in forward modeling \label{tbl:solrat}}
    
    \begin{tabular}{lccc}
\hline
        & \multicolumn{3}{c}{Component} \\
        & (I) & (II) & (III) \\
\hline
        $T$ (K)  & $4600$ & $3200$ & $5400$   \\
        $\Delta\tau_{continuum}$ & $0.3$  & $0.02$ & $0.0001$ \\
        $\Delta\tau_{Fe}$ & $1.1$  & $2$    & $0.09$   \\
        $\Delta\tau_{Ni}$    & $1.2$  & $0.05$ & $0.001$  \\
        $\Delta\tau_{Mn}$    & $3$    & $0.05$ & $0.001$  \\
        $B$ (kG) & $1.9$  & $0$    & $22$    \\
        $\chi_{B}$ & $0$  & $0$    & $0$    \\
        $\theta_{B}$ & $0$  & $0$    & $0$    \\
        Voigt $a$ (Fe) & $5$    & $3$    & $0$      \\
        Voigt $a$ (Mn) & $0.1$  & $0$    & $0$      \\
        Voigt $a$ (Ni) & $0.2$  & $0$    & $0$      \\
        $v_{turb}$ (Fe, km/s)  & $4$    & $0$    & $2.2$    \\
        $v_{turb}$ (Ni, km/s)  & $4.5$  & $0$    & $2.2$    \\
        $v_{turb}$ (Mn, km/s)  & $6$    & $0$    & $2.2$    \\
        $v_{mac}$ (km/s) & $0$    & $0$    & $0.6$    \\
        
\hline
    \end{tabular}
\end{table}

It should be noted that within this simulation, the Voigt profile coefficients $a$ and the turbulent velocities $v_{turb}$ were used as free fitting parameters to reproduce the Lorentzian and non-thermal Gaussian broadening of each of the spectral lines that originate from all relevant line broadening mechanisms. Therefore, these parameters were chosen differently for different spectral lines, and should not to be interpreted as representing only van-der-Waals damping or unresolved sub-telescopic velocities. It should be noted that the parameter set in Table \ref{tbl:solrat} illustrates one physically plausible way to reproduce the observed relative behavior of the three lines, but it should not be interpreted as a unique atmospheric inversion.

Since the Land\'e factor of the Fe I 5434.5 \AA{} line is exactly zero under strict LS coupling, an artificial spin operator scaling factor $\xi=1.014$ (where $\xi=1$ corresponds to LS coupling) was introduced when solving the Paschen-Back eigenvalue problem to mimic the experimental Land\'e factor of $-0.014$ (using the standard notation from \cite{landi2004polarization}):

\begin{eqnarray*}\label{diag}
    &&\langle \beta LSJM | H_{so} + H_{B} | \beta LSJM \rangle =  E_{\beta LS}(J) + \\
    && \qquad\qquad + \mu_{0} BM \left( 1 + \xi \frac{J(J + 1) + S (S + 1) - L(L + 1)}{2J  (J + 1)}\right) ,
\end{eqnarray*}

\begin{eqnarray*}\label{nondiag}
    &&\langle \beta L S\, J-1\, M | H_{\mathrm{so}} + H_B | \beta L S J M \rangle = -\xi\frac{\mu_0 B}{2J} \times\\
    &&\quad \times  \sqrt{\frac{(J+S+L+1)(J-S+L)(J+S-L)(-J+S+L+1)(J^2-M^2)}{(2J+1)(2J-1)}}.
\end{eqnarray*}

\begin{figure}
	\centerline{\includegraphics[width=0.7\textwidth]{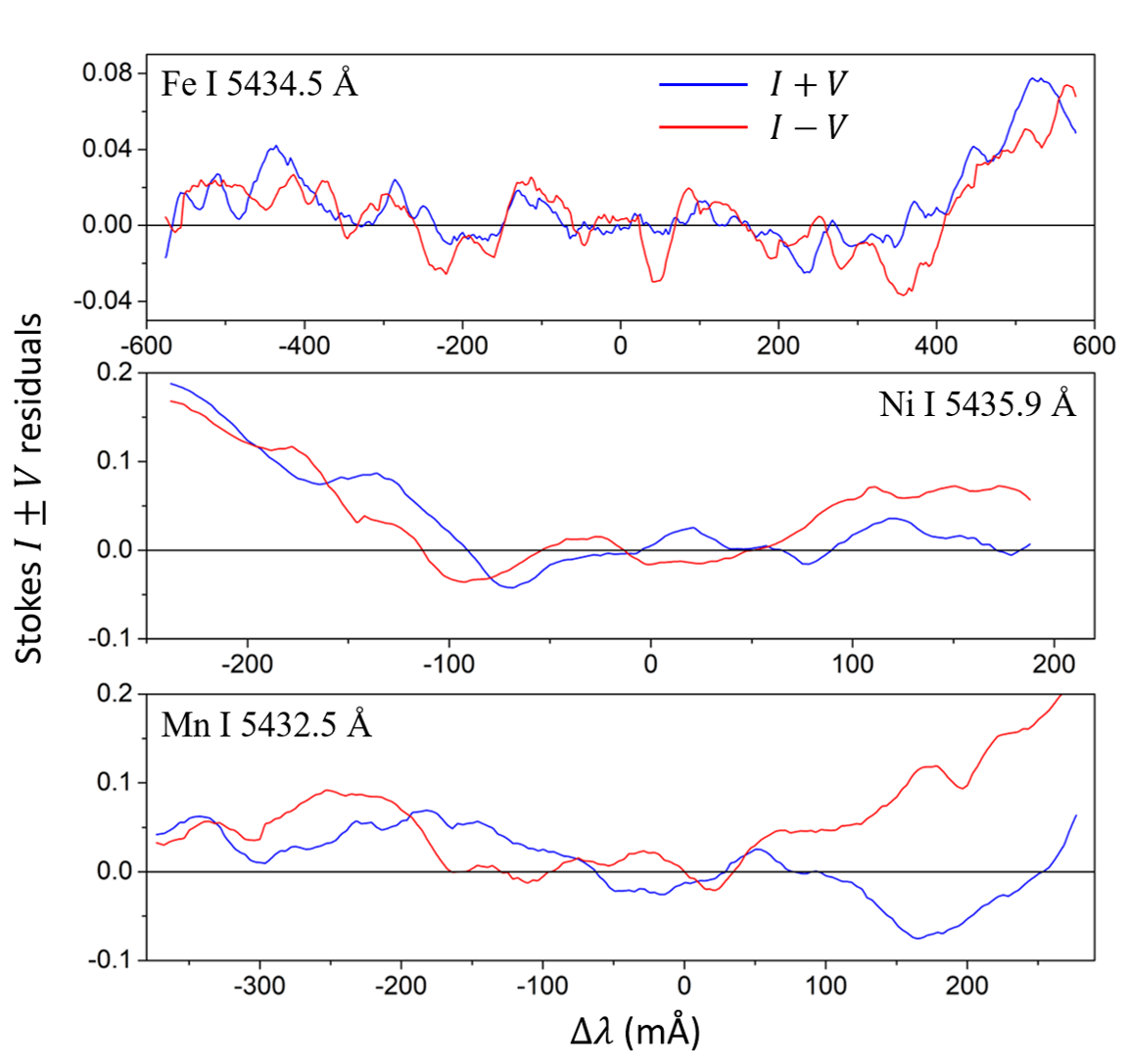}}
\caption{
Difference between the synthetic and observed Stokes $I\pm V$ profiles in Fig. \ref{fig:ipmv}.
}
    \label{fig:resid}
\end{figure}

Figure \ref{fig:resid} shows the difference between observed and synthetic Stokes $I\pm V$ profiles. Since the main purpose of the forward modeling was to illustrate a possible physical mechanism for achieving the observed bisector splitting in the core of the spectral lines, the residuals in Fig. \ref{fig:resid} are naturally lower in the core and increase toward the wings.

\begin{figure}
	\centerline{\includegraphics[width=0.7\textwidth]{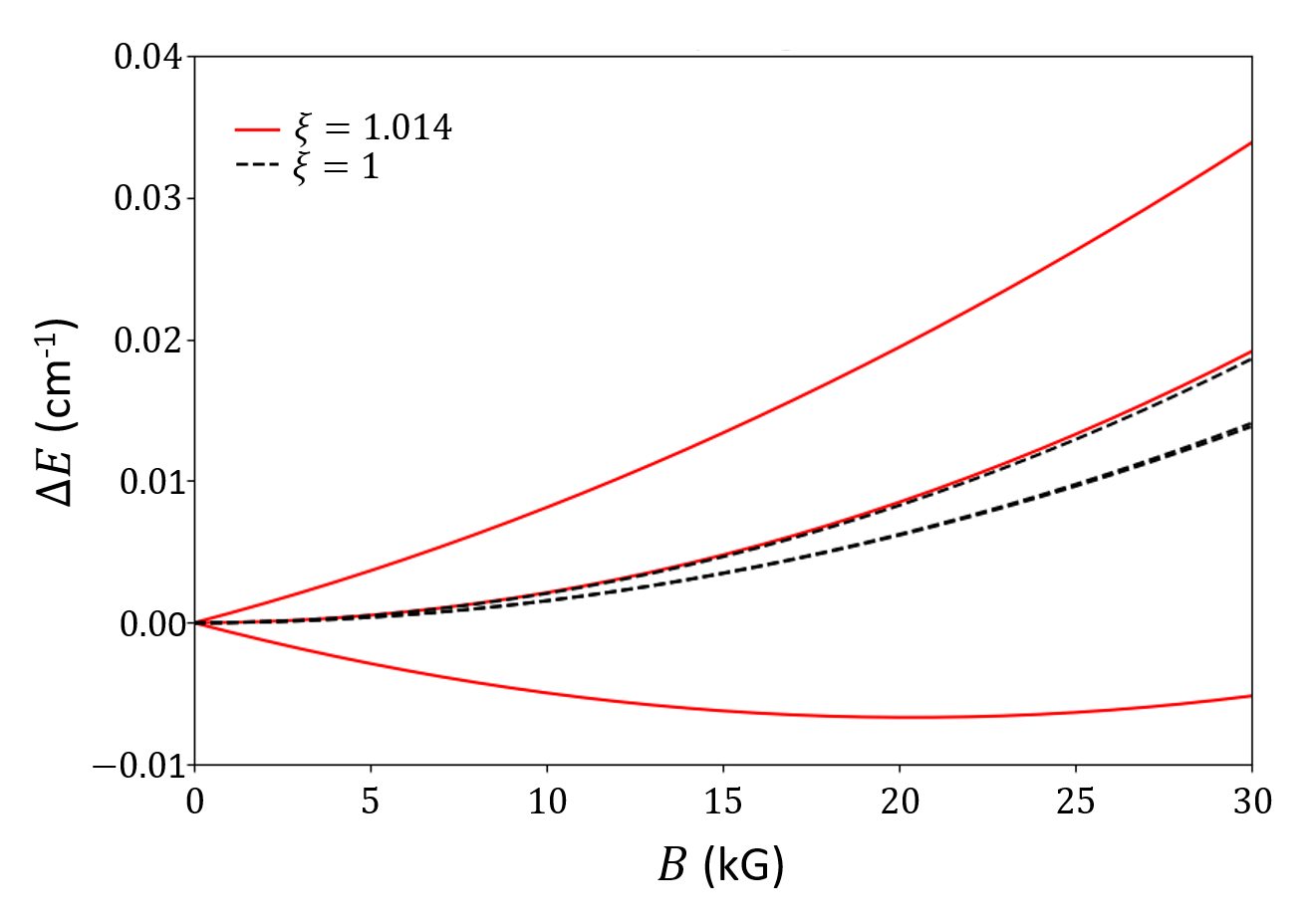}}
\caption{
Magnetic sensitivity of the lower level of the Fe I 5434.5 \AA{} line with (solid) and without (dashed) spin operator scaling. 
}
    \label{fig:magsens}
\end{figure}

Figure \ref{fig:magsens} shows the dependence of the magnetic splitting of the lower level $^{5}F_{1}$ with and without the spin operator scaling. When no spin scaling is applied, the lower level is insensitive to the magnetic field $B \lesssim 5$ kG, since the Land\'e factor of this line is zero under LS coupling. As the field grows, the non-diagonal elements start to have a larger impact and introduce a magnetic splitting by mixing the states with different $J$. When the spin operator is scaled by the factor $\xi=1.014$ to mimic the experimental magnetic sensitivity of the Fe I 5434.5 \AA{} line, the $^{5}F_{1}$ level shows magnetic splitting for arbitrarily low fields. While second-order corrections start to manifest at $B\gtrsim 5$ kG, the overall magnetic sensitivity is still determined mainly by the scaling factor $\xi=1.014$, resulting in the effective Land\'e factor of approximately $-0.014$ across a broad $0$--$30$ kG magnetic field range.

The Mn I 5432.5 \AA{} line results from the transition $z{}^8P_{5/2}\to a{}^{6}S_{5/2}$, which is forbidden under the strict LS coupling due to the selection rule $S=3.5 \not\to S=2.5$. However, since this transition involves only two levels, both of which have $J=5/2$, the radiative transfer equations can be solved by expanding the upper level in the basis of ${}^6P_{5/2}$ and ${}^8P_{5/2}$ by introducing a single parameter $\alpha$: $|\text{upper}\rangle = \sin(\alpha) |{}^{6}P_{5/2}\rangle + \cos(\alpha) |{}^{8}P_{5/2}\rangle$, which effectively results in the radiative transfer equations being multiplied by $\sin^2(\alpha)$ after the selection rules are applied.

It should be noted that forward modeling of the Ni I 5435.9 \AA{} and Mn I 5432.5 \AA{} lines was performed within LS coupling, meaning that their effective Land\'e factors were slightly lower -- $0.500$ and $2.143$, respectively \citep{moore1966solar} -- than the experimental ones used for the calculation of the magnetic parameter $B^*$. However, this difference in the magnetic sensitivity has a minimal impact on the resulting Stokes $I \pm V$ profiles. 

Finally, we note that the forward modeling was performed assuming the heliocentric angle is equal to zero degrees, while the actual heliocentric angle of the sunspot was approximately $32^\circ$. However, since this simulation is performed under the LTE assumption, the non-zero heliocentric angle can be accounted for by treating the optical depths in Table \ref{tbl:solrat} as the line-of-sight optical depths.

\bibliographystyle{unsrtnat}
\bibliography{main}

\end{document}